\documentclass[journal,12pt,onecolumn,draftclsnofoot,]{IEEEtran}%

\usepackage{xcolor,soul,framed} 

\colorlet{shadecolor}{yellow}
\usepackage[pdftex]{graphicx}
\graphicspath{ {photo/} }
\DeclareGraphicsExtensions{.pdf,.jpeg,.png}
\usepackage{xcolor}
\usepackage[linesnumbered,ruled,vlined]{algorithm2e}
\usepackage{amsmath}
\usepackage{array}
\usepackage{mdwmath}
\usepackage{mdwtab}
\usepackage{eqparbox}
\usepackage{url}
\usepackage{subfigure}
\usepackage{amssymb}

\usepackage{algorithmic}

\SetCommentSty{mycommfont}

\SetKwInput{KwInput}{Input}    
\SetKwInput{KwRepeat}{Repeat} 
\SetKwInput{KwOutput}{Output}              
\begin{document}
\bstctlcite{IEEEexample:BSTcontrol}
    \title{\huge Energy Consumption Optimization in RIS-Assisted Cooperative RSMA Cellular Networks}
\author{Shreya Khisa, Mohamad Elhattab, Chadi Assi and Sanaa Sharafeddine \thanks{S. Khisa, M. Elhattab and C. Assi are with Concordia University, Montreal, Quebec, H3G 1M8, Canada (email: shreyakhisa21@gmail.com, mohelhattab@gmail.com, assi@mail.concordia.ca).}
\thanks{Sanaa Sharafeddine is with Lebanese American University Beirut, Beirut 1102 2801, Lebanon (e-mail: sanaa.sharafeddine@lau.edu.lb).}
}
\maketitle

\begin{abstract}
This paper presents a downlink reconfigurable intelligent surface (RIS)-assisted half-duplex (HD) cooperative rate splitting multiple access (C-RSMA) networks. The proposed system model is built up considering one base station (BS), one RIS, and two users. With the goal of minimizing the network energy consumption, a joint framework to optimize the precoding vectors at the BS, common stream split, relaying device transmit power, the time slot allocation, and the passive beamforming at the RIS subject to the power budget constraints at both the BS and the relaying node, the quality of service (QoS) constraints at both users, and common stream rate constraint is proposed. The formulated problem is a non-convex optimization problem due to the high coupling among the optimization variables. To tackle this challenge, an efficient algorithm is presented by invoking the alternating optimization technique, which decomposes the original problem into two sub-problems; namely, sub-problem-1 and sub-problem-2, which are alternatively solved. Specifically, sub-problem-1 is to jointly optimize the precoding vectors, common stream split, and relaying device power. Meanwhile, sub-problem-2 is to optimize the phase shift matrix at the RIS. In order to solve sub-problem-1, an efficient low-complexity solution based on the successive convex approximation (SCA) is proposed. Meanwhile, and with the aid of difference-of-convex rank-one representation and the SCA approach, an efficient solution for the phase shift matrix at the RIS is obtained. The simulation results demonstrate that the proposed RIS-assisted HD C-RSMA achieves a significant gain in minimizing the total energy consumption compared to the RIS-assisted RSMA scheme, RIS-assisted HD cooperative non-orthogonal multiple access (C-NOMA), RIS-assisted NOMA, HD C-RSMA without RIS, and HD C-NOMA without RIS.
\end{abstract}
\vspace{1.5cm}

\IEEEpeerreviewmaketitle

\section{Introduction}
\subsection{Motivation}
"Connecting the unconnected” is an overriding goal of the next generation cellular network (6G) that is driving research to provide seamless and ubiquitous connectivity to every device, given a continued exponential growth of Internet-of-Things (IoT) devices in the next decade. Substantive research programs are being developed on the global scale to shape and establish the vision of 6G including “6G Hubs” in Germany, “6G Flagship” in Finland, Terahertz (THz) communication studies in the USA, “Broadband Communications and New Networks” in China, etc \cite{9693417}.
\par In order to support the different IoT use cases and the new emerging applications requirements such as augmented reality, 4k videos, online gaming, etc., 6G should accommodate a much higher density of connectivity (estimated to be $10^6$ devices per km$^2$), provide 5-10, and 10-100 times of the spectral efficiency and energy efficiency in comparison with the ones that 5G can achieve, 0.9999999 reliability, and support latency in terms of microseconds \cite{9693417}. In order to achieve the aforementioned strict goals, one of the most fundamental issues is to design sophisticated multiple access techniques for the forthcoming wireless networks denoted as next-generation multiple access (NGMA) \cite{9831440}.
\par Rate splitting multiple access has been envisioned as a contender non-orthogonal transmission mechanism for the next-generation wireless networks \cite{9831440}. RSMA is a generalized joint framework of the space division multiple access (SDMA) and the non-orthogonal multiple access (NOMA) and is capable of outperforming both SDMA and NOMA in terms of network energy efficiency, network spectral efficiency, and multiplexing gain \cite{9831440}. Specifically, the main concept of RSMA is to split the user messages into common and private parts wherein the common parts are encoded into common streams, meanwhile, the private parts are encoded into private streams \cite{9831440}. It is worth mentioning that the common streams are required to be decoded by multiple users whereas the private streams are needed to be decoded by the corresponding users \cite{mao2018rate, mao2019rate1}. Accordingly, RSMA provides a versatile interference management feature that allows for partial decoding and partial treatment of interference as noise \cite{mao2020max}. 
\par Note that, in RSMA, the achievable data rate of the common stream is constrained by the rate of the worst user's performance, i.e., the user having the worst channel conditions, which may negatively affect the overall system performance \cite{mao2020max}. As a result, in settings where users experience heterogeneous channel conditions with the transmitter, i.e., the base station (BS), the common stream rate may drop \cite{mao2020max}. In order to tackle this limitation, cooperative communications have been recently incorporated with RSMA to improve the common stream rate, which is known as cooperative RSMA (C-RSMA) \cite{mao2020max}. In particular, since the common stream may carry a portion of each user equipment (UE)'s message, each UE will be able to have prior information about the portion of the messages intended for other UEs. C-RSMA exploits the availability of this prior information about the messages intended for other UE to improve the common stream rate as well as to enhance the reception reliability. To achieve that, the UEs with good channel gains can act as cooperative relays to assist the BS in transmitting the common stream message to the UEs with bad channel conditions. While RSMA/C-RSMA are expected to unleash the potential of 6G networks, another new degree-of-freedom can be added by configuring and controlling the wireless propagation environment, which can be realized through the reconfigurable intelligent surface (RIS).
\par RIS has recently been envisioned as a promising technology to intelligently reconfigure the wireless propagation environment. RIS is composed of a large number of passive low-cost elements, each of which is capable of independently tuning the phase-shift of the impinging radio waves. This technology is envisaged to be deployed as a planar surface on the facades, walls, or ceilings of buildings to reflect the impinging radio waves towards the intended receiver, and hence, can create a virtual line-of-sight (LoS) between the source and the destination. When deployed in wireless systems, RIS has the potential of altering the wireless environment from being highly probabilistic in nature to become more controllable and partially deterministic space referred to as a smart radio environment or \textquotedblleft Wireless 2.0\textquotedblright~\cite{RIS_Open}. Motivated by the above benefits of both C-RSMA and RIS, this paper investigates the potential gains brought by integrating RIS with C-RSMA. Nevertheless, this amalgamation between C-RSMA and RIS can provide a vital paradigm for the forthcoming wireless networks by controlling the wireless propagation through RIS as well as improving the network connectivity as well as network spectral and energy efficiencies through RSMA.
\subsection{State-of-the-art}
\textbf{1) RSMA-empowered wireless networks}: Recently, many works have been devoted to studying and characterizing the potential gains of RSMA-assisted next-generation cellular networks in terms of spectral efficiency, energy efficiency, network coverage, and user fairness \cite{9831440}. The authors in \cite{9195473} studied the joint optimization of beamforming and rate in order to maximize both energy and spectral efficiencies. An optimization problem with the objective of maximizing the minimum achievable rate has been studied in \cite{9145200} for a multi-group multi-cast downlink multiple-input single-output (MISO) system. The authors in \cite{7470942} investigated the performance of RSMA in a massive multiple-input multiple-output (MIMO)-enabled multi-cell scenario to maximize the weighted sum rate. The authors in \cite{7555358} studied the sum-rate maximization problem with the consideration of  partial channel state information (CSI).
\par \textbf{2) C-RSMA-empowered wireless networks}: With the objective of improving the far user performance, and hence, enhancing the overall system performance, C-RSMA with the assistance of half-duplex (HD) and full-duplex (FD) user relaying mode has been recently investigated \cite{8846761,9123680,9627180,9771468}. Specifically, with the objective of maximizing the weighted sum rate, the authors in \cite{8846761} jointly optimized the transmit beamforming at the BS, the transmit power at the near user, the resource allocation, and the common-stream split in a two-user HD C-RSMA network. The authors in \cite{9123680} jointly optimize the BS precoders vectors, the near user's transmit power, the time slot fraction, and the common stream split aiming at maximizing the minimum achievable rate in two-user HD C-RSMA enabled one MISO BS cellular system. On the other hand, the authors in \cite{9627180, 9771468} considered a FD C-RSMA to enhance the system performance in terms of network power consumption \cite{9627180} and user fairness \cite{9771468}.
\par \textbf{3) RIS-assisted RSMA-based wireless networks}: The research on RIS-enabled RSMA-based wireless networks is gaining momentum to enhance different performance metrics \cite{9832618}, such as network energy efficiency \cite{huang2019reconfigurable,yang2020energy}, network spectral efficiency \cite{huang2018achievable, fang2022fully}, and network outage probability \cite{bansal2021rate}. Specifically, in \cite{huang2018achievable}, the authors studied a downlink RIS-assisted wireless communication to maximize the sum rate. The authors in  \cite {huang2019reconfigurable} investigated a RIS-assisted downlink multi-user communication in order to enhance the energy efficiency by considering the QoS constraints of each user. The authors in \cite{yang2020energy} evaluated the energy efficiency maximization in a RIS-assisted RSMA network with the optimization of the phase shift matrices and beamforming vectors of BS.  A framework to leverage the interplay between the RIS and the RSMA is introduced in \cite{bansal2021rate} and a closed form expression for the outage probability of the cell-edge users is derived. In order to maximize the sum rate, the authors in \cite{fang2022fully} proposed a fully connected RIS-assisted RSMA network by optimizing the phase shift matrices of RIS and beamforming vectors. The authors in \cite{9759225} utilized the RSMA technique in a RIS-assisted cloud radio access network (C-RAN) system to improve network energy efficiency. A study on a two-layer RSMA-based multi-RIS system is carried out in \cite{9393472} where each RIS is deployed on a cell-edge boundary to assist the cell-edge users.
\par To the best of our knowledge, all the aforementioned research has only studied the performance of integrating RIS with RSMA technique without adopting the user-relaying cooperation, meanwhile, there is a clear gap in the existing literature on the performance of C-RSMA when the RIS is deployed in the network. In addition, most of the works studying the potential gains of HD C-RSMA in cellular networks devoted their attention to improving the network spectral efficiency as they considered the network spectral efficiency as the key performance indicator for the optimization and design of the wireless systems. However, with the overwhelming rise of mobile devices and traffic data, network energy consumption has severe implications on the economic cost and becomes a major issue in the next-generation wireless networks. Consequently, the energy savings should also attract much attention to a green cellular network, which is the main focus of this paper. 
\subsection{Contributions}
To the best of our knowledge, the integration of RIS with C-RSMA has not been studied in the literature. Moreover, this paper is one of the earliest attempts to investigate the energy consumption minimization problem in RIS-assisted downlink MISO C-RSMA cellular networks. Against the above background and driven by the aforementioned observations, the main contributions of this paper can be summarized as follows.
\begin{itemize}
\item  A MISO downlink RIS-assisted HD C-RSMA system that consists of one multi-antenna BS, one RIS, and two single antenna-based users, i.e., one far user and one near user, is considered. In this model, we allow the near user to assist in the far user transmission by relaying the common stream to this user via a device-to-device (D2D) communication link. In this regard, a framework to optimize the active beamforming at the BS, the power allocation at the near user, the time-slot allocation, the common stream split, and the RIS phase-shift matrix is investigated. This framework is formulated as an optimization problem to minimize the network energy consumption while guaranteeing the QoS requirements for both cellular users and the power constraints of both the BS and the relaying device.
\item Due to the high coupling between the different optimization variables, the formulated total energy consumption minimization problem is a non-convex optimization problem, which is challenging to solve directly. In order to overcome this challenge, we invoke the alternating optimization (AO) method in which the main optimization problem is decomposed into two sub-problems, namely, {sub-problem-1: joint optimization of precoding vectors, common stream split, relaying device power; and sub-problem-2: RIS phase shift matrix optimization}.
\item We solve the sub-problem-1 in an iterative way using successive convex approximation (SCA) by jointly optimizing precoding vectors, common stream split, and relaying device power for a given phase shift matrix at the RIS and a given time slot allocation. Meanwhile, for given precoding vectors, common stream split, time slot allocation, and transmit relaying device power, with matrix lifting and change-of-variables, the passive beamforming optimization sub-problem is reformulated as a rank-one constrained optimization problem. As rank-one constraint is a non-convex term, a difference-of-convex (DC) representation, as well as an efficient SCA algorithm, are considered to solve the reformulated problem in order to obtain a feasible rank-one solution. Furthermore, computational complexity analysis is provided for the overall algorithm. Note that, the optimal time slot allocation is obtained using a exhaustive search over the region from the minimum value to the maximum value of time slot allocation. 
\end{itemize}
Simulations were performed to assess the performance of the proposed RIS-assisted HD C-RSMA framework. The efficacy of the proposed scheme is verified by taking RSMA with RIS, NOMA with RIS, HD C-NOMA with RIS, HD C-RSMA without RIS, and HD C-NOMA without RIS as benchmark schemes.  The simulation results reveal that the proposed RIS-assisted HD C-RSMA scheme can outperform the other five schemes in terms of network energy consumption.
\subsection{Paper Organization and Notations}
The rest of the paper is organized as follows. Section \ref{System Model} presents the network and the transmission models. Section \ref{SINRss} presents the signal-to-interference-plus-noise-ratio (SINR) model and rate analyses. Section \ref{Problem Formulation} discusses the formulated optimization problem and the proposed solution roadmap. Section \ref{Solution Approach} presents the proposed solution approach. Finally, the simulation results and the conclusion are discussed in Sections \ref{Simulation} and \ref{Conclusion}, respectively.
\par Matrices and vectors are denoted by bold-face upper-case and lower-case letters, respectively. For any complex-valued vector $x$, $|x|$ refers to the Euclidean norm of $x$ and $\mathrm{diag}(x)$ denotes the diagonal matrix which has $x$ as diagonal. For any square matrix $\boldsymbol{S}$, $\mathrm{tr}(S)$ refers to as its trace, meanwhile $\boldsymbol{S} \geq 0 $ means that $\boldsymbol{S}$ is a positive semi-definite (PSD) matrix. For any matrix $\boldsymbol{M}$, rank($\boldsymbol{M}$) and $\boldsymbol{M}^H$ refer to as its rank and its conjugate transpose, respectively. Furthermore, the distribution of a circularly symmetric complex Gaussian random variable with a mean $\mu$ and a variance $\sigma^2$ is denoted by $\mathcal{CN}(\mu, \sigma^2)$. In addition, for any matrix, $\boldsymbol{M}$, $||\boldsymbol{M}||_*$ and $||\boldsymbol{M}||_2$ presents the nuclear norm and spectral norm, respectively. Finally, $j$ represents the complex number, $e^{(·)}$ denotes the exponential function, $[h]_n$ is the $n$th entry of $h$, and $\mathcal{R}\{x\}$ is the real part of the complex term $x$.
\section{System Model}
\label{System Model}
\subsection{Network Model}
\begin{figure}
    \centering
    \includegraphics[scale=.4]{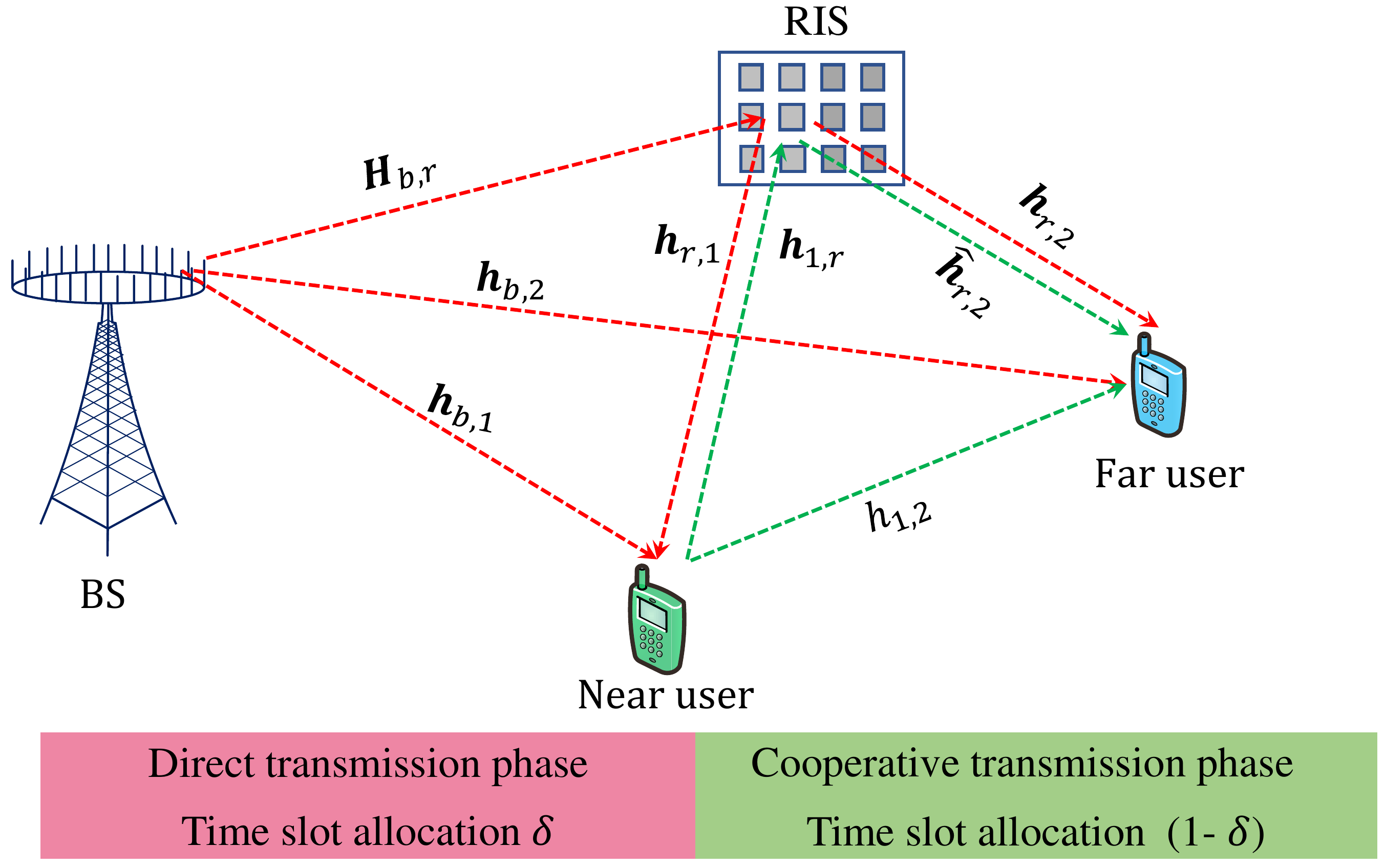}
    \caption{RIS-based C-RSMA System Model}
\label{Fig 1: System Model}
\end{figure}
We consider a downlink transmission in a RIS-assisted C-RSMA system consisting of one BS with $N_t$ number of antennas, two single-antenna users equipment (UEs), and one RIS with $M$ reflecting elements as shown in Fig. 1.  In this model and similar to \cite{zhang2019cooperative}, we assume that user-1, which is near to the BS, has a better channel condition than user-2 which is located far from the BS. Consequently, and based on the distance from BS, user-1 is considered a near user, and user-2 is deemed as a far user \cite{9123680}. By utilizing the principle of C-RSMA and with the assistance of RIS, the BS serves two users. The near user acts as a HD Non-regenerative Decode-and-Forward (NDF) relay to assist the transmission of the common stream from the BS to the far user. Let $\boldsymbol{h}_{b,1} \in \mathbb{C}^{N_t \times 1 }$, $\boldsymbol{h}_{b,2} \in \mathbb{C}^ {N_t \times 1}$, $\boldsymbol{H}_{b,r} \in \mathbb{C}^{N_t \times M}$, $\boldsymbol{h}_{r,1} \in \mathbb{C}^{M \times 1}$, $\boldsymbol{h}_{1,r} \in \mathbb{C}^{M \times 1}$, $\boldsymbol{h}_{r,2}  \in \mathbb{C}^{M \times 1 }$, $\boldsymbol{\hat{h}}_{r,2}  \in \mathbb{C}^{M \times 1 }$ and ${h}_{1,2} \in \mathbb{C}^{1 \times 1}$ be the channel coefficients for the communication links between BS $\rightarrow$ the near user, BS $\rightarrow$ far user, BS $\rightarrow$ RIS, RIS $\rightarrow$ near user, near user $\rightarrow$ RIS, RIS $\rightarrow$ far user in direct phase, RIS $\rightarrow$ far user in cooperative phase, near $\rightarrow$ far user, respectively. Finally, we assume that the channel state information for all these wireless links is perfectly known at the BS.\footnote{To characterize and evaluate the theoretical performance gains and to provide some useful insights into the role of the RIS in improving the performance of C-RSMA, we assume that the CSI of all the wireless communication links is perfectly known. Although it is generally hard to have a perfect CSI, various efforts and studies in the literature have recently developed efficient channel estimation methods for RIS-empowered wireless networks that can be applied in our system model to obtain an accurate CSI. For instance, the authors in \cite{CSI_1} designed an alternating least square technique dependent on the parallel factor framework, which can continuously estimate all the communication links with a low-complexity solution. Two-channel estimation approaches based on compressive sensing and deep learning were developed in RIS-enabled wireless systems.} In the next section, we start to define the received SINRs and the corresponding achievable data rate at the near and far users. 
\subsection {Transmission model} 
In order to proceed, the main operation and transmission phases of the C-RSMA should be first detailed. The transmission model in HD C-RSMA can be executed over two transmission phases, which are denoted as the direct transmission phase and the cooperative transmission phase. As shown in Fig. \ref{Fig 1: System Model}, since the near user adopts HD NDF relaying mode, the two phases may not have equal time allocation. Consequently, we denote the time fractions that are allocated to the direct transmission phase and the cooperative transmission phase as $\delta$ and $(1-\delta)$, respectively. The two transmission phases can be detailed as follows.
\begin{itemize}
    \item {\textbf{Direct transmission phase}}: Following the principle of rate splitting, the BS first splits the message of each UE, i.e, user-1 and user-2, into common and private streams. The common parts are then combined and subsequently encoded into a common stream. Meanwhile, the private streams of the two UEs are independently encoded into private streams. Afterward, the BS transmits the common and private streams to the near and far users. The transmitted signal from the BS also reaches the RIS and is reflected back to the two users. After that, the common stream is decoded by the near user by treating the private streams as interference. Finally, the successive interference cancellation (SIC) process is applied to remove the common stream from the total received signal, and then, each UE decodes its private stream while treating the private stream for the other user as interference.
    \item {\textbf{Cooperative transmission phase}}: By adopting the NDF HD relaying protocol, the near user relays the common stream to the far user through a D2D channel. Consequently, at the far user's side, two signals of the common stream are received, i.e., one coming from the near user transmission and the other one coming from the reflection of the RIS. Finally, at the far user side, those two received signals are combined and the common stream is then decoded.
\end{itemize}
\par Note that the direct transmission phase occurs in the first time slot with a time duration $\delta$; meanwhile, the cooperative transmission phase occurs in the second time slot which has a duration of $(1 - \delta)$. 
\section{SINR Model and Achievable Rate Analysis}
\label{SINRss}
In this section, we discuss the analysis of both the SINRs and the achievable data rates for an HD C-RSMA-based cellular network
assisted by RIS. In this regard, we first discuss the data transmission process by the BS and the near user, and then, the data reception process at the two cellular users. First, at the BS and as we mentioned earlier, the messages for both the near and the far users are split into common $W_{c,1}$ and $W_{c,2}$, respectively, and private parts  $W_{p,1}$ and $W_{p,2}$, respectively. All common parts are encoded together using a common codebook, and then, converted into a common stream $s_c$. Meanwhile, the private parts, i.e. $W_{p,1}$, $W_{p,2}$, are encoded independently and transformed into two private streams $s_1$ and $s_2$, respectively. Note that, the common stream is intended for all users; however, the private streams are intended only for their corresponding users. Based on the above, the signal transmitted by the BS can be expressed as.
\begin{equation}
\boldsymbol{x}=s_c\boldsymbol{p}_c+s_1\boldsymbol{p}_1+s_2\boldsymbol{p}_2. \label{eq:1}
\end{equation}
We denote $\boldsymbol{s}=[s_c,s_1,s_2]^T$ and assume that $\mathbb{E}[\boldsymbol{ss}^H]=\boldsymbol{I}$. Here, the precoding vector adopted by the BS is represented by $\boldsymbol{P}=[\boldsymbol{p}_c,\boldsymbol{p}_1,\boldsymbol{p}_2]$. Consequently, the received signal at the near user in the first time slot can be written as
\begin{equation}
    y_1 = (\boldsymbol{h}_{b,1}^H + \boldsymbol{h}_{r,1}^H \boldsymbol{\Theta}^{[1]}\boldsymbol{H}^H_{b,r})\boldsymbol{x} + \omega_1, \label{eq:2}
\end{equation}
where $\omega_1$ represents the additive white Gaussian noise with $\mathcal{C N}\left(0, \sigma_1^2\right)$ at near user in the first time slot. Moreover, $\boldsymbol{\Theta}^{[1]}$ = diag$\left(\phi_1^{[1]},\phi_2^{[1]},... ,\phi_M^{[1]}\right)$ is the phase shift matrix of the RIS, where $\forall m \in [1,M], \phi_m^{[1]} = e^{j\theta_m^{[1]}}$ such that $\theta_m^{[1]} \in [0,2\pi]$ denotes the phase shift of the $m$-th reflective element in the first time slot. Consequently, the received SINR and the corresponding achievable data rate at the near user to decode the common stream at the first time slot can be given by,
\begin{align}
\mathrm{SINR}_{c,1}^{[1]}\left(\boldsymbol{\theta}^{[1]},\boldsymbol{P} \right) &= \frac{|(\boldsymbol{h}_{b,1}^H + \boldsymbol{h}_{r,1}^H\boldsymbol{\Theta}^{[1]}\boldsymbol{H}^H_{b,r})\boldsymbol{p}_c|^2}{\sum_{k=1}^{2}|(\boldsymbol{h}_{b,1}^H+\boldsymbol{h}_{r,1}^H\boldsymbol{\Theta}^{[1]}\boldsymbol{H}^H_{b,r})\boldsymbol{p}_k|^2+\sigma_{1}^{2}}, \label{eq:3}    \\
    R_{c,1}^{[1]} \left(\boldsymbol{\theta}^{[1]},\boldsymbol{P} \right) &=\delta\log_2{\left(1+\mathrm{SINR}_{c,1}^{[1]}\right)}, \label{eq:4}
\end{align}
where $\boldsymbol{\theta}^{[1]}=[\theta_1^{[1]},\theta_2^{[1]},...,\theta_M^{[1]}]$.
Meanwhile, the SINR and the corresponding achievable data rate at the near user to decode its private stream at the first time slot can be expressed as.
\begin{align}
\mathrm{SINR}_{p,1}^{[1]}\left(\boldsymbol{\theta}^{[1]},\boldsymbol{P} \right)&=\frac{|(\boldsymbol{h}_{b,1}^H+\boldsymbol{h}_{r,1}^H\boldsymbol{\Theta}^{[1]}\boldsymbol{H}^H_{b,r})\boldsymbol{p}_1|^2}{|(\boldsymbol{h}_{b,1}^H+\boldsymbol{h}_{r,1}^H\boldsymbol{\Theta}^{[1]}\boldsymbol{H}^H_{b,r})\boldsymbol{p}_2|^2+\sigma_1^{2}}, \label{eq:5} \\
R_{p,1}^{[1]}\left(\boldsymbol{\theta}^{[1]},\mathbf{P} \right) & = \delta\log_2{\left(1+\mathrm{SINR}_{p,1}^{[1]}\right)}. \label{eq:6}
\end{align}
On the other hand, the received signal at the far user at the first time slot can be expressed by,
\begin{equation}
    y_2 =  (\boldsymbol{h}_{b,2}^H + \boldsymbol{h}_{r,2}^H \boldsymbol{\Theta}^{[1]}\boldsymbol{H}^H_{b,r})\boldsymbol{x}+\omega_2^{[1]}, \label{eq:7}
\end{equation}
where $\omega_2^{[1]}$ denotes the additive white Gaussian noise with $\mathcal{CN}\left(0, \sigma_2^{2}\right)$ at the far user in time slot 1. Therefore, the SINR expression and the achievable data rate at the far user to decode the common stream at time slot 1 can be given by,
\begin{align} \mathrm{SINR}_{c,2}^{[1]}\left(\boldsymbol{\theta}^{[1]},\boldsymbol{P} \right) & = \frac{|(\boldsymbol{h}_{b,2}^H+\boldsymbol{h}_{r,2}^H\boldsymbol{\Theta}^{[1]}\boldsymbol{H}^H_{b,r})\boldsymbol{p}_c|^2}{\sum_{k=1}^{2}|(\boldsymbol{h}_{b,2}^H+\boldsymbol{h}_{r,2}^H\boldsymbol{\Theta}^{[1]}\boldsymbol{H}^H_{b,r})\boldsymbol{p}_k|^2+\sigma_2^{2}}, \label{eq:8} \\
R_{c,2}^{[1]}\left(\boldsymbol{\theta}^{[1]},\boldsymbol{P}\right) &=\delta\log_2{\left(1+\mathrm{SINR}_{c,2}^{[1]}\right)}. \label{eq:9}
\end{align}
Moreover, the SINR expression and the achievable rate at the far user to decode its private stream in the first time slot can be expressed as, 
\begin{equation}    \mathrm{SINR}_{p,2}^{[1]}\left(\boldsymbol{\theta}^{[1]},\boldsymbol{P} \right)=\frac{|\boldsymbol{h}_{b,2}^H + \boldsymbol{h}_{r,2}^H \boldsymbol{\boldsymbol{\Theta}}^{[1]} \boldsymbol{H}^H_{b,r})\boldsymbol{p}_2|^2}{|(\boldsymbol{h}_{b,2}^H+ \boldsymbol{h}_{r,2}^H\boldsymbol{\Theta}^{[1]}\boldsymbol{H}^H_{b,r})\boldsymbol{p}_1|^2+\sigma_{2}^{2}}, \label{eq:10}
\end{equation}
\begin{equation}
R_{p,2}^{[1]}\left(\boldsymbol{\theta}^{[1]},\boldsymbol{P}\right)=\delta\log_2{\left(1+\mathrm{SINR}_{p,2}^{[1]}\right)}. \label{eq:11}
\end{equation}
After we characterize the transmission model, the received signal at both UEs, the received SINRs, and the corresponding achievable rate in the first time slot due to the transmission of the BS, we now direct our attention toward analyzing the network model in the second time slot.  At time slot 2, the near user starts to transmit the common stream to the far user as mentioned in the cooperative transmission phase. Hence, the received signal at the far user in time slot 2 can be expressed as.
\begin{equation}
    y_2=(h_{1,2}+\boldsymbol{\hat{h}}_{r,2}^H\boldsymbol{\Theta}^{[2]}_{HD}\boldsymbol{h}_{1,r})\sqrt{P_d}s_c+\omega_2^{[2]},\label{eq:12}
\end{equation}
where $\omega_2^{[2]}$ denotes the additive white Gaussian noise with $\mathcal{CN}\left(0, \sigma_2^{2}\right)$ at far user in time slot 2. Furthermore, $\boldsymbol{\Theta}^{[2]}$= diag
$(\phi^{[2]}_1,\dots, \phi^{[2]}_M)$ is the phase shift matrix of the RIS at time slot 2 such that $\forall m \in [1,M], \phi^{[2]}_m$= $e^{j\theta_m^{[2]}}$. In order to decode the common stream at time slot 2, the SINR and the achievable rate at the far user can be given by
\begin{align}
\mathrm{SINR}_{c,2}^{[2]} & =\frac{|({h}_{1,2}+\boldsymbol{\hat{h}}_{r,2}^H\boldsymbol{\Theta}^{[2]}\boldsymbol{h}_{1,r})|^2P_{\rm d}}{\sigma_{2}^{2}},\label{eq:13} \\
    R_{c,2}^{[2]}\left(\boldsymbol{\theta}^{[2]},P_{\rm d} \right) & =(1-\delta)\log_2{\left(1+\mathrm{SINR}_{c,2}^{[2]}\right)}. \label{eq:14}
\end{align}
Since the near user adopts NDF relaying mode, the common stream at the far user can be expressed by \cite{gunduz2007opportunistic, amin2012opportunistic},
\begin{equation}
R_{c,2}^{\mathrm{NDF}}\left(\boldsymbol{\theta}^{[1]},\boldsymbol{\theta}^{[2]},\boldsymbol{P}, P_{\rm d} \right) = R_{c,2}^{[1]}\left(\boldsymbol{\theta}^{[1]},\boldsymbol{P}\right)+ R_{c,2}^{[2]}\left(\boldsymbol{\theta}^{[2]},P_{\rm d}\right). \label{eq:15}
\end{equation}
Finally, the achievable data rate to decode the common stream for both users can be expressed by
\begin{equation}
R_{c}=\min{\left(R_{c,1}^{[1]}\left(\boldsymbol{\theta}^{[1]},\boldsymbol{P}\right),R_{c,2}^{NDF}\left(\boldsymbol{\theta}^{[1]},\boldsymbol{\theta}^{[2]},\boldsymbol{P}, P_{\rm d} \right) \right)}\label{eq:16}.
\end{equation}
{Specifically, $R_c$ guarantees the successful decoding of the common stream at both users. Since $R_c$ is distributed among the two users, hence, for the common stream $s_c$, $R_c$ should satisfy  $R_c = \sum_{k \in K}C_k$, where $\forall k \in \{1,2\}, C_k$ represents the portion of the common stream that is allocated to UE $k$. After successful decoding and SIC process, $s_c$ is removed from the total received signal, and the private streams are decoded by their corresponding users. It is worth mentioning that the joint optimization of the common stream vector $\boldsymbol{c}=[C_k],\forall k \in \{1,2\}$ is needed to achieve the maximum achievable rate of the worst-case scenario \cite{9123680}.}
\section{Problem Formulation and Solution Roadmap}
\label{Problem Formulation}
\subsection{Problem Formulation}
In this paper, we investigate the joint optimization of the precoding vectors at the BS, i.e., $\boldsymbol{P}$, the common stream split, i.e., $\boldsymbol{c}$, the time slot allocation, i.e., $\delta$, the D2D relaying power, i.e., $P_{\rm d}$, and the phase shift matrix at the RIS in the first and the second time slots, i.e., $\boldsymbol{\Theta}^{[1]}, \boldsymbol{\Theta}^{[2]}$, with the objective of minimizing the total energy consumption while guaranteeing the power budget constraints at both the BS and the near user, the required QoS in terms of the minimum achievable data rate, the common stream rate, and the RIS phase-shift unit-modulus constraint. Accordingly, the network energy consumption minimization problem for RIS-assisted HD C-RSMA can be formulated as follows. 
\allowdisplaybreaks
\begin{subequations}
\label{prob:P1}
\begin{flalign}
\centering
 \mathcal{P}_1: &\min_{\substack{\boldsymbol{P}, \,\boldsymbol{c},\,P_{\rm d},{\delta},\\ \boldsymbol{\theta}^{[1]}, \boldsymbol{\theta}^{[2]} }}\delta(||\boldsymbol{p}_1||^{ 2}+||\boldsymbol{p}_2||^2+||\boldsymbol{p}_c||^2)+(1-\delta)P_{\rm d},\:  \label{eqn:17} \\
 &\text{s.t.} \quad||\boldsymbol{p}_1||^{ 2}+||\boldsymbol{p}_2||^2+||\boldsymbol{p}_c||^2 \le P_{\rm BS},\label{eq:18}\\
    &\quad\quad 0 \le P_{\rm d}\le P_{\rm D2D},\,\,  \label{eq:19}\\
     &\quad\quad C_k\ge 0,\,\, \forall k \in \{1,2\},\label{eq:20}\\
     &\quad \quad C_k+R_{p,k}^{[1]}\left(\boldsymbol{\theta}^{[1]},\mathbf{P} \right) \ge R_{th,k},\,\, \forall k \in \{1,2\},  \label{eq:22}\\
 &\quad \quad C_1+C_2\le R_{c}\left(\boldsymbol{\theta}^{[1]},\boldsymbol{\theta}^{[2]},\boldsymbol{P}, P_{\rm d} \right)  ,\, \label{eq:24}\\
     &\quad\quad  |\phi_m^{[1]}|=1,\,\, \forall m \in [1,M], \label{eq:25}\\
      &\quad\quad |\phi_m^{[2]}|=1,\,\, \forall m \in [1, M], \label{eq:26}
\end{flalign}
\end{subequations}
where \eqref{eq:18} and \eqref{eq:19} refer to the power budget at the BS and the relaying device, respectively, where $P_{\rm D2D}$ denotes the maximum transmit power at the relaying device. \eqref{eq:22} represents the QoS constraints for UE $k$, where $R_{th,k}$ represents the required data rate threshold at UE $k$. \par In order to minimize the network energy consumption, the precoding vectors at the BS, the time slot allocation, the common stream vector split, the transmit power at the relaying user, and the passive beamforming at the RIS should be jointly optimized. Because of the coexistence and the high coupling of the optimization variables, both the objective function and the considered constraints of the optimization problem $\mathcal{P}_1$ are non-convex. In other words, it can be easily observed
that the formulated optimization problem is a non-convex optimization problem that is difficult to directly solve. As a result, it is necessary to divide problem $\mathcal{P}_1$ into some tractable sub-problems which can be solved alternately and separately over multiple iterations. In doing so, we resort to the AO method to solve the original problem $\mathcal{P}_1$ in an efficient manner which will be presented in the following part.
\subsection{Solution Roadmap}
Due to its intractability, and for a given value of time slot allocation, i.e., $\delta$, problem $\mathcal{P}_1$ is decomposed into two sub-problems, namely, joint precoding vectors, power allocation at the relaying node, and common stream split optimization sub-problem (referred to as sub-problem-1), and a passive beamforming optimization sub-problem, which is referred to as sub-problem-2. It is worth mentioning that the two sub-problems are non-convex optimization problems, and with the aid of the AO technique, the two sub-problems are solved in an alternating way \cite{AO_1, AO_2}. Precisely, we optimize the time slot allocation, $\delta \in (0,1]$ through an exhaustive search method. In other words, we obtain the network energy consumption value for each given $\delta$, and then, we select the value of $\delta$ that can achieve the minimum network energy consumption. Note that, for each value of $\delta$ and with the aid of the AO approach, the two sub-problems are solved as follows. At the first iteration, a random phase shift matrix at the RIS is generated and fed to sub-problem-1. After that, the resulting problem is solved and the optimized precoding vectors at the BS, the transmit power at the near user, and the common stream split vector are input to the second sub-problem, i.e., sub-problem-2. Then, the derived passive beamforming sub-problem is solved and the optimized phase shift matrix is obtained, which will be then input to the first sub-problem at the second iteration. Note that, the same steps of the first iteration are repeated. This procedure, which is denoted as the iterative AO method, will keep running until convergence. Toward that end, the formulated sub-problems are shown below.
\subsection{Sub-problem-1: Joint Optimization of  Active Beamforming, Common Stream Split Vector, and Near User Transmit Power}
In this section, for a given value of the phase-shift matrix in each time period, i.e., $\boldsymbol{\Theta}^{[1]}$ and $\boldsymbol{\Theta}^{[2]}$, we study the power allocation policy, which includes the precoding vectors and the near user transmit power and the common stream split vector. This problem can be formulated as follows.\footnote{Throughout the rest of the paper, the terms ``active beamforming'' and ``precoding vectors'' will be used interchangeably. In addition, ``passive beamforming'' and ``phase shift matrix'' will be used interchangeably.} 
\allowdisplaybreaks
\begin{subequations}
\label{eq:P2}
\begin{flalign}
\centering
 \mathcal{P}_2\quad: &\min_{\boldsymbol{P},\,\boldsymbol{c}, \, P_{\rm d}} \quad \delta(||\boldsymbol{p}_1||^{ 2}+||\boldsymbol{p}_2||^2+||\boldsymbol{p}_c||^2)+(1-\delta)P_{\rm d},\:  \label{eq:27} \\
 & \eqref{eq:18}-\eqref{eq:24}.
\end{flalign}
\end{subequations}
One can see that problem $\mathcal{P}_2$ is neither concave nor
quasi-concave due to the non-convex rate constraints, and hence, the optimal solution for problem $\mathcal{P}_2$ is challenging to obtain in practice. In order to tackle this issue, we provide an efficient solution, which is based on the SCA method, in Section \ref{Solution Approach}-A.
\subsection{Sub-problem-2: Passive Beamforming at the RIS}
In this sub-problem, and for given values of the precoding vectors at the BS, the transmit power at the near user, and the common stream split vector, we need to obtain the phase shift matrix of the RIS in the first and the second time slots. Accordingly, the passive beamforming optimization problem can be presented as follows.
\allowdisplaybreaks
\begin{subequations}
\label{eq:P3}
\begin{flalign}
\centering
\mathcal{P}_3\quad:\quad & \rm find \quad \quad \boldsymbol{\theta}^{[1]}, \boldsymbol{\theta}^{[2]},\\ \label{eqn:28} 
 & \eqref{eq:22}-\eqref{eq:26}.
\end{flalign}
\end{subequations}
One can see that problem $\mathcal{P}_3$ is a feasibility check problem (finding the phase shift for each reflecting element such that the required QoS are guaranteed). However, problem $\mathcal{P}_3$ is a non-convex optimization problem due to the non-convex unit modulus constraints \eqref{eq:25} and \eqref{eq:26}. In order to resolve this challenge, we provide an efficient solution for this problem in Section \ref{Solution Approach}-B.
\section{Proposed Solution Approach}
\label{Solution Approach}
We present the proposed solution approach for both sub-problem-1 and sub-problem-2. We start by solving sub-problem-1, and then, we provide the solution approach for sub-problem-2.
\subsection{Joint Optimization of Precoding Vectors, Common Stream Split, and Relaying Device Power: Solution Approach}
For sub-problem-1, we assume that phase shift matrices $\boldsymbol{\Theta}^{[1]}$ and $\boldsymbol{\Theta}^{[2]}$ are fixed. Now, for the sake of simplicity, we denote the following terms as follows:
\begin{align}
\boldsymbol{h}_{k}^H & =\boldsymbol{h}_{b,k}^H+\boldsymbol{h}_{r,k}^H\boldsymbol{\Theta}^{[1]}\boldsymbol{H}^H_{b,r},\label{eq:29} \\
\hat{h}_{1,2} & = {h}_{1,2} + \boldsymbol{\hat{h}}_{r,2}^H\boldsymbol{\Theta}^{[2]}\boldsymbol{h}_{1,r}, \\
\eta & = \delta(||\boldsymbol{p}_1||^{ 2}+||\boldsymbol{p}_2||^2+||\boldsymbol{p}_c||^2)+(1-\delta)P_{\rm d}.\label{eq:30}
\end{align}
In order to tackle the non-convex constraints in \eqref{eq:22},
we introduce slack variables  $\boldsymbol{\gamma}_p =[\gamma_{p,k}], \forall k \in \{1,2\}$,  where they represent the SINRs for the private streams. Consequently, \eqref{eq:22} can be reformulated as follows.
\begin{align}
 C_k+ \delta \log_2(1+\gamma_{p,k}) &\ge R_{th,k}, \quad \forall k \in \{1,2\},\label{eq:31}\\
 \frac{|\boldsymbol{h}_k^H\boldsymbol{p}_k|^2}{\sum_{j \in \mathcal{K}, j \neq k}|\boldsymbol{h}_{k}^{H}\boldsymbol{p}_j|^2+\sigma_k^{2}} & \geq \gamma_{p,k}, \quad \forall k \in \{1,2\}. \label{eq:32}
\end{align}

  However, \eqref{eq:32} is still non-convex, and hence, a new slack variable $\boldsymbol{\beta}_p =[\beta_{p,k}]$, $\forall k \in \{1,2\}$ is introduced that represents the interference-plus-noise term for the private streams. Therefore, \eqref{eq:32} can be transformed as follows.
\begin{align}
 \frac{|\boldsymbol{h}_k^H\boldsymbol{p}_k|^2}{\beta_{p,k}} &\geq \gamma_{p,k}, \quad \forall k \in \{1,2\},\label{eq:33}\\
  \beta_{p,k} &\ge \sum_{j \in \mathcal{K}, j \neq k}|\boldsymbol{h}_k^H\boldsymbol{p}_j|^2+\sigma_k^{2}, \quad \forall k \in \{1,2\}.
  \label{eq:34}
  \end{align}
Similarly, for constraint \eqref{eq:24},  we introduce slack variables $\boldsymbol{\gamma}_c =[\gamma_{c,k}]$, $\forall k = \{1,2\}$, and $\boldsymbol{\beta}_c =[\beta_{c,k}]$, $\forall k = \{1,2\}$,  where $\boldsymbol{\gamma}_c$ represents the SINRs of common stream and $\boldsymbol{\beta}_c$ represents the interference-plus-noise term for the common stream. Therefore, \eqref{eq:24} can be rewritten as follows. 
\begin{subequations}
\begin{flalign}
  &C_1+C_2 \le \delta \log_2(1+\gamma_{c,1}), \label{eq:35}\\
   &C_1+C_2 \le \delta \log_2(1+\gamma_{c,2})+(1-\delta)
   \log_2{(1+\frac{|{\hat{h}_{1,2}}|^2P_{d}}{\sigma_{2}^2})}, \label{eq:36}\\
  &\frac{|\boldsymbol{h}_k^H\boldsymbol{p}_c|^2}{\beta_{c,k}} \geq \gamma_{c, k}, \forall k \in \mathcal{K}\label{eq:37},\\
  &\beta_{c,k} \ge \sum_{j \in \mathcal{K}}|\boldsymbol{h}_k^H \boldsymbol{p}_{j}|^{2}+\sigma_k^{2}.\label{eq:38}
  \end{flalign}
  \end{subequations}
However, non-convex terms still exist in \eqref{eq:33}  and \eqref{eq:37}. Note that, the terms $\frac{|\boldsymbol{h}_k^H\boldsymbol{p}_{k}|^{2}}{\beta_{p,k}}$ and $\frac{|\boldsymbol{h}_k^H\boldsymbol{p}_{c}|^{2}}{\beta_{c,k}}$ follow a  generic form as, $f(u, v)=\frac{|v|^{2}}{u}, \forall v \in \mathbb{C}, \forall u \in \mathbb{R}^{+} $. Toward this end, using the lower bounded concave approximation, we approximate function $f(u,v)$ on point $u^{(n)}, v^{(n)}$ in order to solve the non-convex terms in an iterative manner. The  non-convex terms in  \eqref{eq:33}  and \eqref{eq:37} can be approximated with the following generic equation below
\cite{mao2018energy}:
 \begin{equation}
 f(u, v) \geq F\left(u, v ;u^{(n)}, v^{(n)}\right)=\frac{2\mathcal{R}\left\{v^{(n)*} v\right\}}{u^{(n)}}-\frac{\left|v^{(n)}\right|^{2}}{\left(u^{(n)}\right)^{2}} u, \label{eq:39}
 \end{equation}
 where $\mathcal{R}$ represents the real number.
 Finally, using the above approximations, $\mathcal{P}_2$ can be transformed as follows.
\begin{subequations}
\begin{flalign}
\centering
\mathcal{P}_4\quad:\quad &\min_{\mathbf{P},\,\mathbf{c}, \, P_{\rm d}, \, \,\boldsymbol{\gamma}_{p},\, \boldsymbol{\gamma}_{c}, \, \boldsymbol{\beta}_{p}, \, \boldsymbol{\beta}_{c}} \quad \quad \eta\: \label{eq:43} \\
&\frac{2\mathcal{R}({\boldsymbol{{p}}_k^n}^H\boldsymbol{h}_k\boldsymbol{h}_k^H\boldsymbol{p}_k)}{{\beta}_{p,k}^n}-\frac{|\boldsymbol{h}_k^H\boldsymbol{p}_k^n|^2\beta_{p,k}}{(\beta_{p,k}^{n})^2}
\geq \gamma_{p,k}, \label{eq:44} \\
&\frac{2\mathcal{R}({\boldsymbol{p}_c^n}^H\boldsymbol{h}_k\boldsymbol{h}_k^H\boldsymbol{p}_c)}{\beta_{c,k}^n}-\frac{|\boldsymbol{h}_k^H\boldsymbol{p}_c^{n}|^2{\beta_{c,k}}}{(\beta_{c,k}^n)^2} \ge \gamma_{c,k},\label{eq:45}\\
 &  \eqref{eq:31}, \eqref{eq:34}, \eqref{eq:35}, \eqref{eq:36}, \eqref{eq:38}.
\end{flalign}
\end{subequations}
One can see that problem $\mathcal{P}_4$ is a convex second-order cone program (SOCP) that can be solved using any convex optimization solver such as YALIMP or CVX. The SCA-based algorithm to solve $\mathcal{P}_4$ is detailed in \textbf{Algorithm 1}.
\begin{algorithm}[!t]
\caption{SCA-based algorithm for Problem ($P_4$)}\label{alg:one}
\KwInput{Time slot allocation $\delta$, tolerance $\epsilon_1$, $\boldsymbol{\theta}^{[1]}$, and $\boldsymbol{\theta}^{[1]}$.}
\textbf{Initialize}: \rm $\boldsymbol{P}^0$, $\eta^0$,  $\boldsymbol{\beta}_{c}^0$, $\boldsymbol{\beta}_{p}^0$\;
$n=0$\;
\While{$|\eta^n$ - $\eta^{n-1}| > \epsilon_1$ }{
$n=n+1$\;  
solve $\mathcal{P}_4$ using  $\boldsymbol{P}^{n-1}, \eta^{n-1},
  \boldsymbol{\beta}_{p}^{n-1},
  \boldsymbol{\beta}_{c}^{n-1}$\;
  Find the optimization variables $\boldsymbol{P}^*, \eta^*,
  \boldsymbol{\beta}_{p}^{*},            \boldsymbol{\beta}_{c}^{*}$\;   
  Update $\boldsymbol{P}^n$ $\leftarrow \boldsymbol{P}^*$, ${\eta}^n$, $\leftarrow \eta^*$, $\boldsymbol{\beta}_{p}^n \leftarrow \boldsymbol{\beta}_{p}^*$, 
 $\boldsymbol{\beta}_{c}^n \leftarrow \boldsymbol{\beta}_{c}^*$
}
\end{algorithm}
\subsection{Sub-problem-2: Phase Shift Optimization: Solution Approach}
In this section, we discuss the phase shift optimization of the RIS during both the direct and the cooperative transmission phases. It should be noted that the main objective of the RIS is to provide additional paths to construct a stronger combined channel gain at the intended receiver. Consequently, the best channel gain at the far user, in the second time slot, can be achieved when the reflected signals from the RIS's meta-atoms can be constructively added at the far user and be aligned with the direct D2D link from the near user to the far user. As a result, the optimal phase-shift coefficient in the cooperative transmission phase can be obtained as follows.
\begin{align}
    \theta_m^{[2]} & = \arg(h_{1,2})-\arg\left([\mathbf{h}_{1,r}]_m\mathbf{[\hat{h}}_{r,2}]_m\right), \label{eq:43} \\ 
\mathrm{SINR}_{c,2}^{[2]} & =\frac{\left(|h_{1,2}|+\sum_{m=1}^{M}|[\mathbf{h}_{1,r}]_m\mathbf{[\hat{h}}_{r,2}]_m|\right)^2 P_{d}}{\sigma_{2}^{2}}. \label{eq:44}
\end{align}
After obtaining $\boldsymbol{\theta}^{[2]}$, one can see that problem $\mathcal{P}_3$ is still non-convex. This is because there still exists the unit modulus constraint in \eqref{eq:25} and \eqref{eq:26}. To tackle this challenge and to determine the phase shift matrix in the first time slot, we reformulate $\mathcal{P}_3$ as a rank-one constrained optimization problem via matrix lifting and change-of-variables. Afterward, the rank-one constraint is converted to a DC problem. In the end, an effective SCA approach is adopted to solve the rank-one constraint optimization in an iterative manner.
\subsection{Rank-one constrained optimization problem }
We define $\boldsymbol{v} \triangleq \left[\phi^{[1]}_1,\phi^{[1]}_2,...\phi^{[1]}_M\right]^H$ where $\forall m = 1,\dots,M$, $\boldsymbol{h}_{b,k}^H \boldsymbol{p}_j=b_{k,j}$. Now, by applying a change-of-variables 
$\boldsymbol{h}_{r,k}^H\boldsymbol{\Theta}^{[1]}\boldsymbol{H}^H_{b,r} \boldsymbol{p}_j= \boldsymbol{v}^H\boldsymbol{a}_{k,j}$ where $\boldsymbol{a}_{k,j} =$ diag$(\boldsymbol{h}^H_{r,k})\boldsymbol{H}_{b,r}^H\boldsymbol{p}_j$. Now, we introduce an auxiliary variable $t$, problem $\mathcal{P}_3$ can be rewritten as follows:
\allowdisplaybreaks
\begin{subequations}
\begin{align}
 \mathcal{P}_5: \quad& \mathrm{Find} \quad \boldsymbol{V},\label{eq:46}\\ 
 &|{b}_{k,k}|^2 +  \mathrm{tr} \left(\boldsymbol{Q}_{k,k}\boldsymbol{V}\right) \ge  \mu_{k}\sum_{j \neq k} \mathrm{tr} \left(\boldsymbol{Q}_{k,j}\boldsymbol{V}\right)+ \mu_{c,1}\left(\sum_{j \neq k}|{b}_{k,j}|^2+   \sigma_{k}^{2}\right), \quad \forall k \in \{1,2\} \label{eq:47}\\
&|{b}_{k,c}|^2 + \mathrm{tr}\left(\boldsymbol{Q}_{k,j}\boldsymbol{V}\right)\ge \mu_{c,1}\sum_{k=1}^{2}\mathrm{tr}\left(\boldsymbol{Q}_{k,j}\boldsymbol{V}\right) +
\mu_{c,1}\left(\sum_{k=1}^{2}|\boldsymbol{b}_{k,j}|^2+\sigma_{k}^{2}\right), \quad \forall k \in \{1,2\}\label{eq:48}\\
&|{b}_{k,c}|^2 + \mathrm{tr}\left(\boldsymbol{Q}_{k,j}\boldsymbol{V}\right)\ge\mu_{c,2}\sum_{k=1}^{2} \mathrm{tr}\left(\boldsymbol{Q}_{k,j}\boldsymbol{V}\right)+\mu_{c,2}\left(\sum_{k=1}^{2}|{b}_{k,j}|^2+\sigma_{k}^{[1]^2}\right) + R_{c,2}^{[2]}, \quad  \forall k \in \{1, 2\}\label{eq:49}\\
&[\boldsymbol{V}]_{m,m} = 1, \quad m=1, \dots, M+1, \label{eq:50}\\
&\boldsymbol{V} \succeq 0, \label{eq:51}\\
&\mathrm{rank} (\boldsymbol{V})=1.\label{eq:52}
 \end{align}
 \end{subequations}
where
\begin{equation}
\begin{split}
 \boldsymbol{Q}_{k,j}=\begin{bmatrix} \boldsymbol{a}_{k,j} \boldsymbol{a}_{k,j}^H &  \boldsymbol{a}_{k,j} {b}_{k,j}^H \\
 \boldsymbol{a}_{k,j}^H{b}_{k,j}  & 0 \end{bmatrix}, \quad & \boldsymbol{\bar{v}}=\begin{bmatrix} \boldsymbol{v} \\ t   \end{bmatrix},
 \end{split}
\end{equation}
and, 
\begin{align}
\mu_k & =2^\frac{(R_{th,k}-C_k)}{\delta}-1, \quad   \mu_{c,1} =2^\frac{(C_1+C_2)}{\delta}-1, \quad \mathrm{and} \quad \mu_{c,2}=2^\frac{(C_1+C_2-R_{c,2})}{\delta}-1.
\end{align}
It should be noted that $\boldsymbol{V} \triangleq \boldsymbol{\bar{v}\bar{v}}^H$ is required to satisfy rank-one constraint. Due to the non-convex nature of the rank-one constraint, it is still hard to solve $\mathcal{P}_5$ in a straightforward manner. One popular approach considered in the literature to resolve the rank-one constraint is to adopt a semi-definite relaxation (SDR). By dropping the rank-one constraint, the obtained optimization problem ends up with convex semidefinite programming (SDP) which can be efficiently solved by traditional convex optimization solvers such as YALIMP or CVX. However, applying the SDR approach to solve a feasibility check problem during the AO approach does not guarantee the feasibility of the given solution \cite{yang2020federated, yu2021irs}. In specific, the rank-one solution achieved by the Gaussian randomization (GR) method is not guaranteed to maintain the QoS constraints, which causes an early stopping of the AO iterations \cite{yang2020federated, yu2021irs}.
\subsection{DC Representation for Rank-one Constraint}
To deal with the drawbacks resulting from eliminating the rank-one constraint and with the aim to enhance the performance degradation in the SDR technique, we develop a DC format of rank-one constraint. This format ensures a feasible solution for the phase-shift optimization problem. It should be noted that for a PSD matrix $\boldsymbol{V} \in \mathbb{C}^{(M+1) \times (M+1)}$, rank one constraint means that $\sigma_1(\boldsymbol{V})\ge 0$ and $\sigma_m(\boldsymbol{V})= 0$ for all $m \in [2, M+1]$, where for $m \in [1,M+1]$, $\sigma_m(\boldsymbol{V})$ denotes the $m$-the largest eigenvalue of $\boldsymbol{V}$. Accordingly, we can represent the rank-one constraint as follows.
\begin{equation}
\rm {rank} (\boldsymbol{V}) = 1 \Leftrightarrow ||\boldsymbol{V}||_*-||\boldsymbol{V}||_2 = 0,\label{eq:53}
\end{equation}
where $||\boldsymbol{V}||_*= \sum_{m=1}^{M} \sigma_m (\boldsymbol{V})$ and $||\boldsymbol{V}||_2= \sigma_1(\boldsymbol{V})$ represent
the nuclear and the spectral norms of $\boldsymbol{V}$, respectively. Consequently, problem $\mathcal{P}_5$ can be reformulated as follows:
\allowdisplaybreaks
\begin{subequations}
\label{eq:P7}
\begin{flalign}
\centering
\mathcal{P}_6\quad:\quad &
\min_{\boldsymbol{V}} ||\boldsymbol{V}||_*-||\boldsymbol{V}||_2,\label{eq:54}\\
& \eqref{eq:47}-\eqref{eq:51}.
\end{flalign}
\end{subequations}
It can be seen from $\mathcal{P}_6$ that a rank-one feasible solution can be obtained when the objective of \eqref{eq:P7} reaches zero. Note that, since $||\boldsymbol{V}||_2$ is a convex function, the $\mathcal{P}_6$ is still non-convex
optimization problem. Hence, we can apply SCA to solve it in an iterative manner. In particular, by linearizing the convex term, all that remains is to solve the following optimization problem.
\allowdisplaybreaks
\begin{subequations}
\label{eq:P8}
\begin{flalign}
\centering
\mathcal{P}_7\quad:\quad &
\min_{\boldsymbol{V}}|\boldsymbol{V}\|_{*}-\left\langle\partial_{\boldsymbol{V}^{[r-1]}}\left\|\boldsymbol{V}^{[r-1]}\right\|_{2}, \boldsymbol{V}\right\rangle,\label{eq:55}\\
& \eqref{eq:47}-\eqref{eq:51}.
\end{flalign}
\end{subequations}
where $\boldsymbol{V}^{[r-1]}$ is the obtained solution at iteration $r-1$, $\partial_{\boldsymbol{V}^{[r-1]}}\left\|\boldsymbol{V}^{[r-1]}\right\|_{2}$ is the sub-gradient of the spectral norm at point $\boldsymbol{V}^{[r-1]}$. $\langle., .\rangle$ denotes the inner product that is given by $\langle\boldsymbol{V}, \boldsymbol{Z}\rangle=\Re\left(\operatorname{tr}\left(\boldsymbol{V}^{H} \boldsymbol{Z}\right)\right)$. We can evaluate $\partial_{\boldsymbol{V}^{[r-1]}}\left\|\boldsymbol{V}^{[r-1]}\right\|_{2}$ as $\boldsymbol{v}_1\boldsymbol{v}_1^H$ where $\boldsymbol{v}_1$ is the eigenvector corresponding to the largest singular value $\sigma_1(\boldsymbol{V})$. Given an initial value of $\boldsymbol{V}^{[0]}$ and by iteratively solving problem $\mathcal{P}_7$ until the objective reaches zero, we can guarantee an exact rank-one solution for the phase shift matrix. One possible stopping criterion for problem $\mathcal{P}_7$ is given by $||\boldsymbol{V}||_*-||\boldsymbol{V}||_2 \leq \zeta_{\rm DC}$, where $\zeta_{\rm DC} > 0$ is a sufficiently small constant \cite{yang2020federated}. The overall algorithm which is referred to as AO-based RIS-assisted HD C-RSMA is described in \textbf{Algorithm 2}. Particularly, \textbf{Algorithm 2} optimizes precoding vectors, common stream split, relaying device power and phase shift matrices in an alternating manner by performing exhaustive search over time slot allocation.
\begin{algorithm}[!t]
\DontPrintSemicolon
  \KwInput{$P_{\rm D2D}$, $P_{\rm BS}$, $\sigma_1^{2}$, $\sigma_2^{2}$, $\delta_{min}=0.1$, $\delta_{max}=1$, $\mathrm{SINR}_{c,2}^{[2]}$}.
 \textbf{Initialization}: Phase shift $\boldsymbol{\Theta}^{[1]}$, iteration $i=1$, convergence $\epsilon$, $\zeta_{\rm DC}$, $t = 0$, and the max number of iterations $J_1$ \;
  \For{$\delta_{min}$ to $\delta_{max}$}{
  $t := t + 1$\;
  \textbf{repeat}\;
  For a given $\boldsymbol{\Theta}^{[1]}$ and using \textbf{Algorithm 1}, find the precoding vectors at the BS $\boldsymbol{P}$, the common stream split $\boldsymbol{c}$, and the near user transmit power $P_{\rm d}$. \;
  $r=1$.\;
  \While{$\mathrm{objective~of}$ $\mathcal{P}_8$ $\ge$ $\epsilon$~$\mathrm{or}$~$r$ $\le J_2$}{
  Calculate the sub-gradient $\partial_{\boldsymbol{V}^{[r-1]}}\left\|\boldsymbol{V}^{[r-1]}\right\|_{2}$ and obtain $\boldsymbol{V}^r$ by solving (38).\;
  Update $r := r+1$\;
  }
  Utilizing Cholesky decomposition get $\boldsymbol{\bar{v}}^{[i]}$, where $\boldsymbol{V}^r=\boldsymbol{\bar{v}}^{[i]}\left(\boldsymbol{\bar{v}}^{[i]}\right)^H$.\;
  Get the phase shift matrix $\boldsymbol{\Theta}^{[1]} = diag\left(\left(\boldsymbol{v}^{[i]}\right)^H\right)$ where $\boldsymbol{v}^{[i]} =\left(\boldsymbol{\bar{v}}_{[1:M]}/\boldsymbol{\bar{v}}_{[M+1]}\right)$\;
  Update $i := i+1$\;
\textbf{Until} objective value at $\mathcal{P}_4$ $\le$ $\epsilon$ or the number of iterations = $J_1$.\;
Calculate the network energy consumption and store it in $\mathcal{Q}(t)$.\;}
Network energy consumption = $\min_t \mathcal{Q}(t)$.
\caption{AO-based RIS-assisted HD C-RSMA}
\end{algorithm}
\subsection{Computational Complexity analysis}
In order to measure the computational complexity of \textbf{Algorithm 2}, we need to analyze the complexity of the exhaustive search, sub-problem-1, and sub-problem-2. Note that, we consider the exhaustive search over the interval of $(0,1]$ with a step size of value 0.1. In each iteration of the exhaustive search, \textbf{Algorithm 2} is solved. 
For sub-problem-1, the complexity burden stems from solving the process of SCA.  $\mathcal{P}_4$ is a second-order cone program with a complexity of
$(S^2_1 S_2)$, where $S_1 = (5+N_t){K}+N_t+2$ is the total number of
variables and $S_2 = 7K + 2$ is the total number of constraints \cite{mao2020max}, where $K$ is the number of users, i.e., $K=2$. Therefore, the  complexity of the \textbf{Algorithm 1} can be obtained by $O(N_t^2{K}^{3.5})$ \cite{mao2020max}. On the other hand, the computational complexity for sub-problem-2 can be obtained as follows. Since the phase shift optimization problem is SDP, interior points methods are widely utilized to solve the SDP problem \cite{polik2010interior}. According to the Theorem 3.12 of \cite{polik2010interior}, the computational complexity of the SDP problem, with $l$ SDP constraints which have $b \times b$ PSD matrix, can be calculated as $\mathcal{O}(\sqrt b \log (1/\epsilon_3)) (lb^3+l^2b^2+l^3)$, where $\epsilon_3$ represents the solution accuracy. In $\mathcal{P}_7$, there are $b=M+1$ and $l=4K+M+1$. Therefore, the approximate complexity of the phase shift optimization sub-problem can be obtained by $\mathcal{O}(J_2\log (1/\epsilon_3)M^{4.5})$. $J_2$ is the maximum number of iterations that are required until the objective goes below the adopted threshold. Therefore, the approximated total complexity of the overall algorithm can be given by $\mathcal{O}(N_{itr} J_1{(N_t^2{K}^{3.5} + J_2\log(1/\epsilon_3) M^{4.5}}))$, where $N_{itr}$ represents the number of steps that is required for the exhaustive search. 
\section{Simulation results and discussions}
\label{Simulation}
In this section, we evaluate the performance of the proposed RIS-empowered HD C-RSMA cellular networks. In order to evaluate the efficacy of the proposed scheme, we compare it with five benchmark schemes under various system parameters by varying the rate threshold at the far UE, the number of RIS elements, and the location of the RIS. These benchmark schemes can be detailed as follows.\footnote{It is worth mentioning that, according to the work in the literature, the HD C-RSMA achieves a higher performance gain than the general RSMA  \cite{mao2020max}\cite{zhang2019cooperative}\cite{9771468}. Based on this observation, we exclude the general RSMA from the benchmark schemes.}
\begin{itemize}
    \item \textit{RSMA with RIS} \cite{9145189}: A multi-antenna-based general RIS-assisted RSMA framework without user cooperation is applied in the considered two-user scenario. This scheme can be obtained from the proposed framework by excluding the near user power transmit from the optimization and setting the value of $P_{\rm d} = 0$.
    \item \textit{NOMA with RIS} \cite{9197675}: A multi-antenna-based general RIS-assisted non-orthogonal multiple access (NOMA) scheme without user relaying cooperation is assumed for the adopted two-users case. 
    \item \textit{HD C-NOMA with RIS} \cite{9586734}: A multi-antenna-based RIS-assisted HD Cooperative NOMA (C-NOMA) scheme is considered. During the cooperative transmission phase and by utilizing the D2D link, the near user relays the information of the far user using HD user relaying mode. 
    \item \textit{HD C-RSMA without RIS} \cite{9123680}:
    This scheme follows a similar mechanism to our proposed scheme without utilizing the RIS. Consequently, it is only required to solve sub-problem-1 by eliminating the effect of the RIS from the channel gain. 
    \item \textit{HD C-NOMA without RIS} \cite{7117391}: This scheme is a typical HD C-NOMA scheme without the assistance of the RIS.
\end{itemize}
\subsection{Simulation settings}
Our simulation setting includes one BS, one RIS, one near user, and one far user, and their locations are given in a three-dimensional coordinate system as follows: $(0, 10 \mathrm{m}, 0), (80 \mathrm{m}, 10\mathrm{m}, 0)$, $(40\mathrm{m}, 0, 0)$, and $(80 \mathrm{m}, 0, 0)$, respectively. For the communication links, we consider both small-scale and large-scale fading. The large-scale fading follows a distance-based path-loss model such as $PL{(d_x)} = \rho_0 (d_x/d_0)^{-\eta_x}$, where $\rho_0$ represents the path-loss exponent at a reference distance $d_0$, $\eta_x$ represents the path-loss exponent, and $d_x$ represents the distance between two nodes. It is assumed that line-of-sight (LoS) communication links exist between the BS and the RIS $(\boldsymbol{H}_{b,r})$ and the communication link between the RIS and the far user $(\boldsymbol{h}_{r,2})$. Therefore, they experience small-scale fading that follows a Rician fading. Now, we can express the channel coefficients for Rician fading as
\begin{align}
    \boldsymbol{h}_x= \sqrt{PL({d_x}})\left(\sqrt{\left(\frac{1}{1+\lambda_x}\right)}m_x+\sqrt{\left(\frac{1}{1+\lambda_x}\right)}\hat{m}_x\right),
\end{align} 
where $\lambda_x$ represents the Rician factor, $m_x$ represents the LoS channel, and $\hat{m}_x$ represents the Non-LoS (NLoS) channel that follows a Rayleigh distribution with zero mean and variance one. The following communication links follow Rayleigh distribution with zero mean and unit variance: $\boldsymbol{h}_{b,1}$, $\boldsymbol{h}_{b,2}$, $\boldsymbol{h}_{r,1}$. Hence, the channel coefficients for these links can be expressed as $\boldsymbol{h}_{b,1}=\boldsymbol{q}_{b,1}\sqrt{PL({d_{b,1}})}$, $\boldsymbol{h}_{b,2} = \boldsymbol{q}_{b,2}\sqrt{PL({d_{b,2}})}$, $\boldsymbol{h}_{r,1} = \boldsymbol{q}_{r,1} \sqrt{PL({d_{r,1}})}$. Here, $\boldsymbol{q}_x$ represents the Rayleigh fading with zero mean and unit variance and $PL({d_x})$ represents the large-scale path-loss component for the communication link $x$.

\begin{table}
\centering
\caption{Simulation parameters}
\begin{tabular}{ |l|c|c| }
 \hline
  \textbf{Parameter} & \textbf{Symbol} & \textbf{Value} \\
 \hline
 Path-loss exponent for the BS-RIS and RIS-far user& $\eta_{br}$, $\eta_{rf}$   & 2.2  \\
 \hline
 Path-loss exponent for the BS-far user & $\eta_{bf}$ &   4 \\
 \hline
Path-loss exponent for the near user-RIS and near user-far user & $\eta_{nr}$, $\eta_{nf}$ & 3 \\
 \hline
Path-loss exponent for the BS-near user & $\eta_{bn}$ & 3.5 \\
 \hline
 Noise power at near and far user & $\sigma_{1}^{2}$, $\sigma_{2}^{2}$ & -120 dB  \\
 \hline
 Path loss exponent at reference distance of 1m & $\rho_o$& -30 dB \\
 \hline
 BS power budget & $P_{\rm BS}$& 53 dBm \\
 \hline
 Relaying device power budget & $P_{\rm d}$& 30 dBm  \\
 \hline
\end{tabular}
\end{table}

\begin{figure}
    \centering
    \includegraphics [scale=.35]{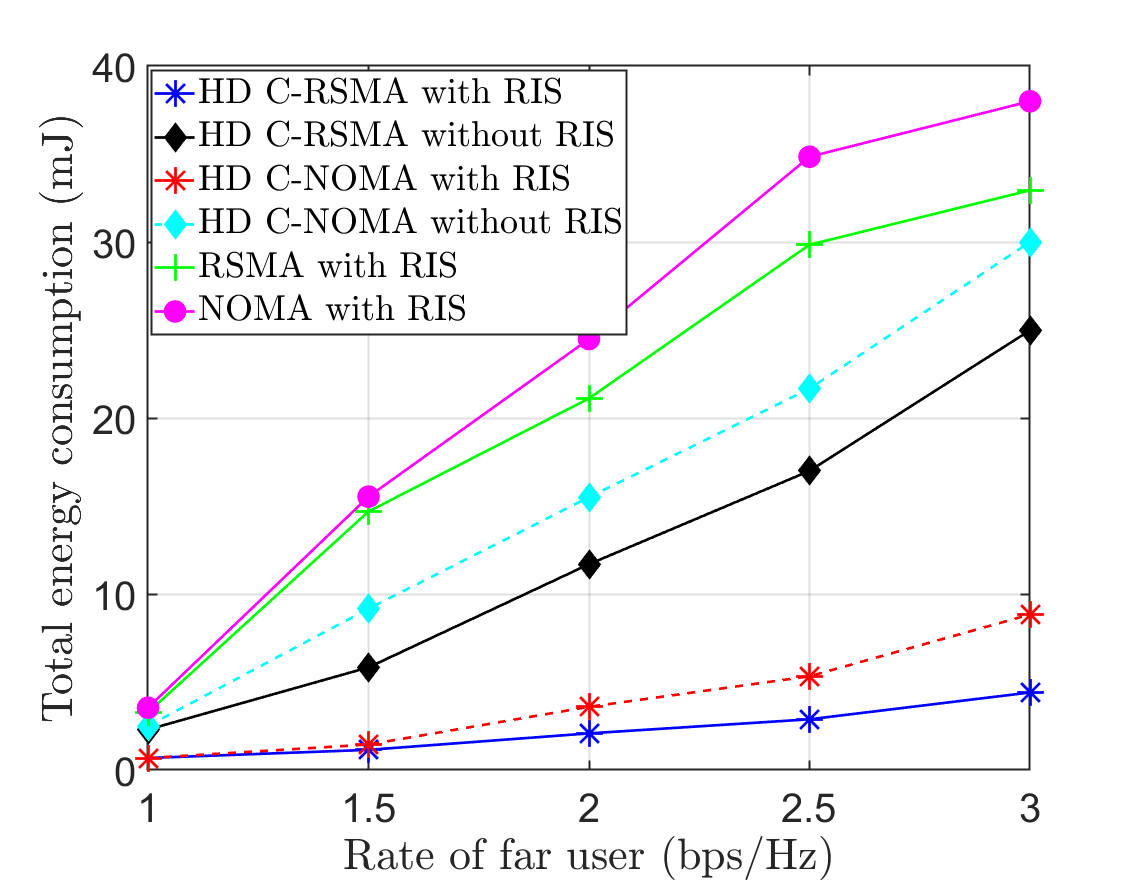}
    \caption{Energy consumption vs rate threshold of the far user when $M=40$.}
\label{Fig 3: rate}
\end{figure}
\subsection{The Impact of The Far User QoS Requirements}
Fig. \ref{Fig 3: rate} demonstrates the total energy consumption  for the proposed scheme and the five baseline schemes vs the data rate threshold of the far user when $M=40$. It can be seen that the total energy consumption of the proposed scheme is lower than the HD C-RSMA without the RIS scheme. The main reason behind this is that the great improvement in the channel gains due to the RIS and the high spectral efficiency provided by the C-RSMA both allow the BS as well as the near user to reduce their transmit power, which leads to lower energy consumption. In the first time slot, the systems with cooperation, the BS transmits the signal and both users benefit from the transmission of the BS as well as the reflection of the RIS due to that transmission. In the second time slot, the near user relays the common stream to the far user via the D2D link. Note that the near user transmission also reaches the RIS and, with the proper configuration of the RIS, the RIS reflects that signal to the far user in a constructive way. Therefore, the far user benefits from both RIS and the D2D cooperation. Furthermore, it can also be seen from the figure that when the far user data rate is near 1 bps/Hz, HD C-RSMA with RIS and HD C-NOMA with RIS have almost similar performance. This observation is also valid for the case between RSMA with RIS and NOMA with RIS, and between HD C-RSMA without RIS and HD C-NOMA without RIS. However, as the data rate threshold of the far user increases, the gap between the base schemes also increases. This is due to the fact that RSMA has the capability to lower energy consumption by treating the multi-user interference either as interference or noise. Meanwhile, in NOMA, multi-user interference is considered pure interference, and every user is impelled to decode the message of all other users. On the contrary, for the case of general RSMA and NOMA with RIS where cooperation does not exist, the BS needs to transmit with more power to achieve the required data rate which results in high energy consumption. 
\subsection{The Impact of The Number of the RIS Elements}
\begin{figure}
    \centering
    \includegraphics[scale=.35]{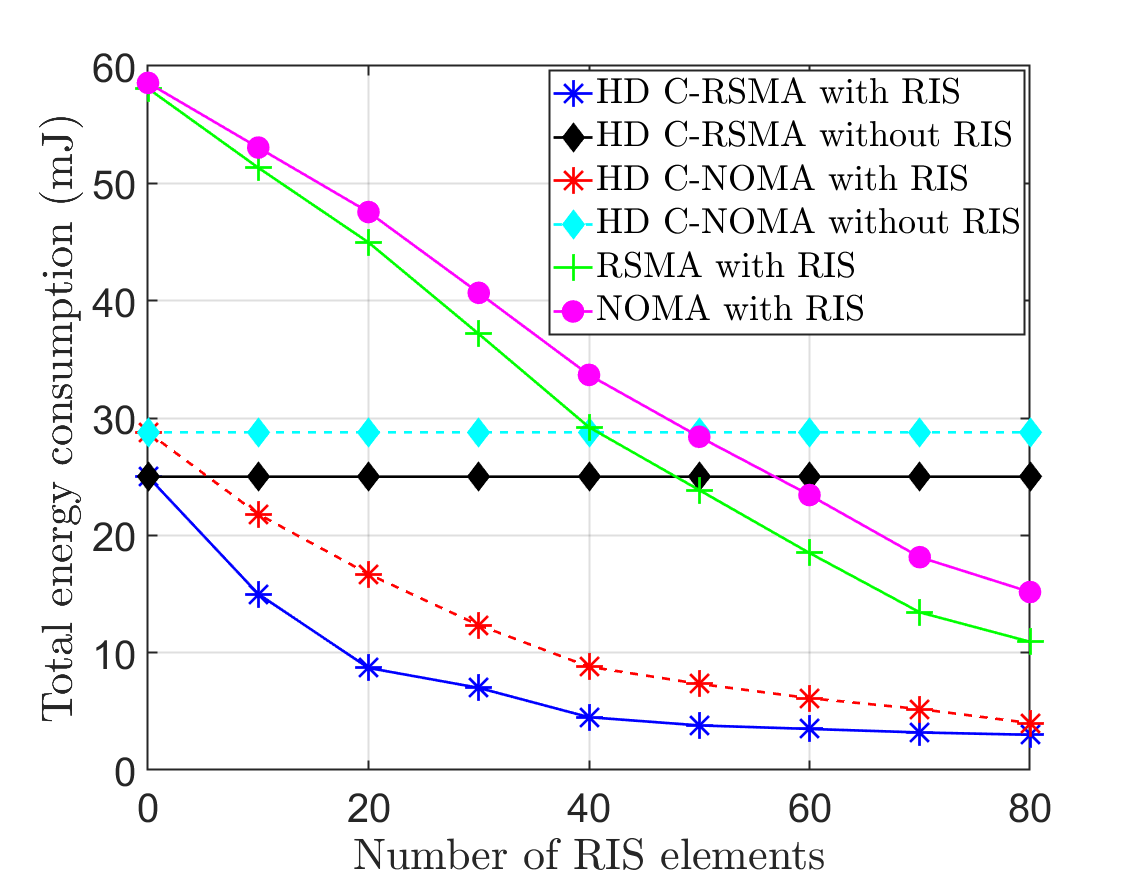}
    \caption{Energy consumption vs Number of RIS elements when $R_{th,1}= 1$ bps/Hz, $R_{th,2}= 3$ bps/Hz.}
\label{Fig 4: RIS elements}
\end{figure}
Fig. \ref{Fig 4: RIS elements} depicts the impact of the number of the RIS elements vs the total energy consumption of our proposed scheme as well as the five baseline schemes. It can be seen from Fig. \ref{Fig 4: RIS elements} that as the number of RIS elements increases, the total energy consumption decreases for all schemes utilizing the RIS. This observation is expected due to the fact that a large number of RIS elements can obtain higher combined channel gains, which results in higher passive array gains. Higher RIS elements provide additional degrees of freedom to the transmitted signal during the direct and cooperative phase. It is also visible from the figure that as the number of RIS elements increases the gap between the HD C-RSMA and HD C-NOMA decreases. It is due to the fact that both schemes can leverage the benefits of the high number of RIS elements and user cooperation. However, the proposed HD C-RSMA with RIS still defeats the HD C-NOMA when the number of RIS elements is large for both schemes. Furthermore, it can also be observed from the figure that a large number of RIS elements allows the non-cooperative schemes (RSMA with RIS and NOMA with RIS) to beat the cooperative ones without RIS (HD C-RSMA without RIS and HD C-NOMA without RIS) at a moderate and high number of RIS elements.   However, the proposed HD C-RSMA can significantly outperform other schemes which illustrate the superiority of integrating RIS with the C-RSMA. Finally, one can see that HD C-NOMA requires a high number of RIS elements to achieve the same performance as the HD C-RSMA.
\subsection{Impact of the RIS location}
\begin{figure*}
    \centering
    \subfigure[]{\includegraphics[scale=0.25]{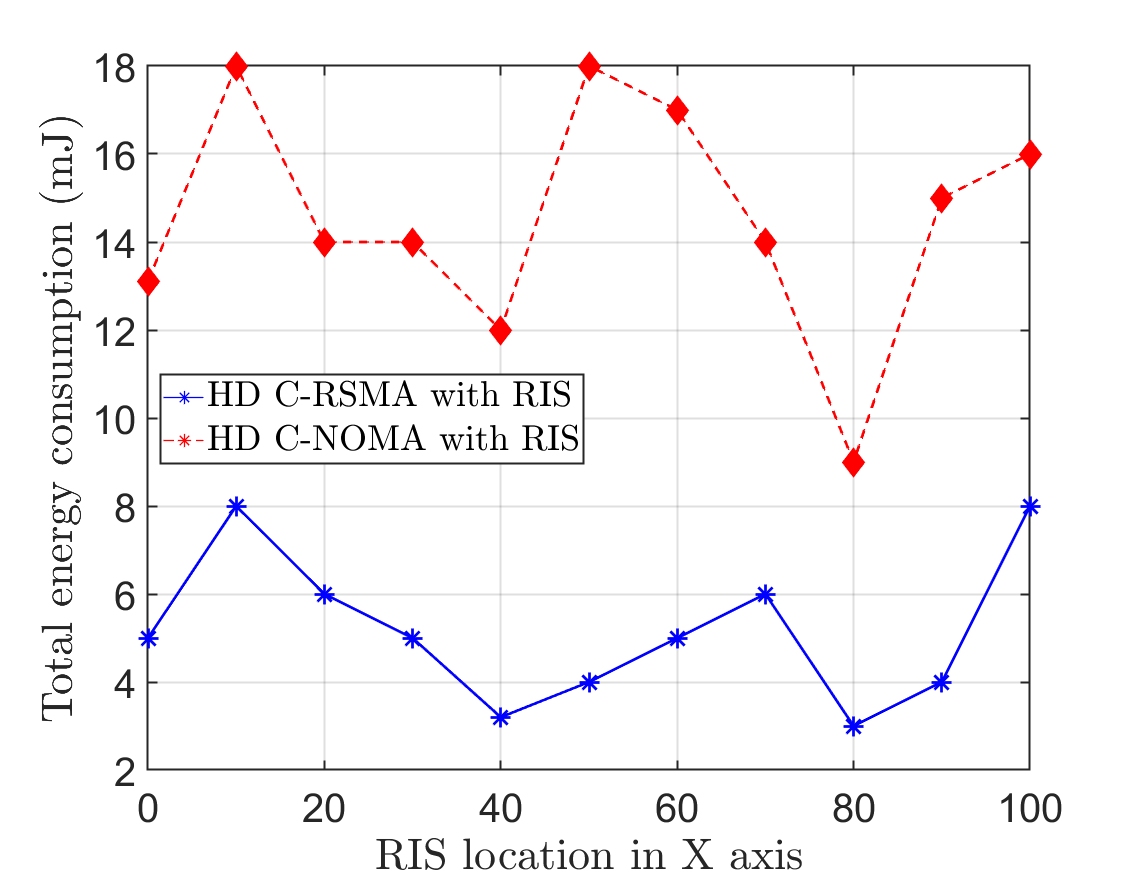}\label{Fig 5: RIS loc crsma}}
    \subfigure[]{\includegraphics[scale=0.25]{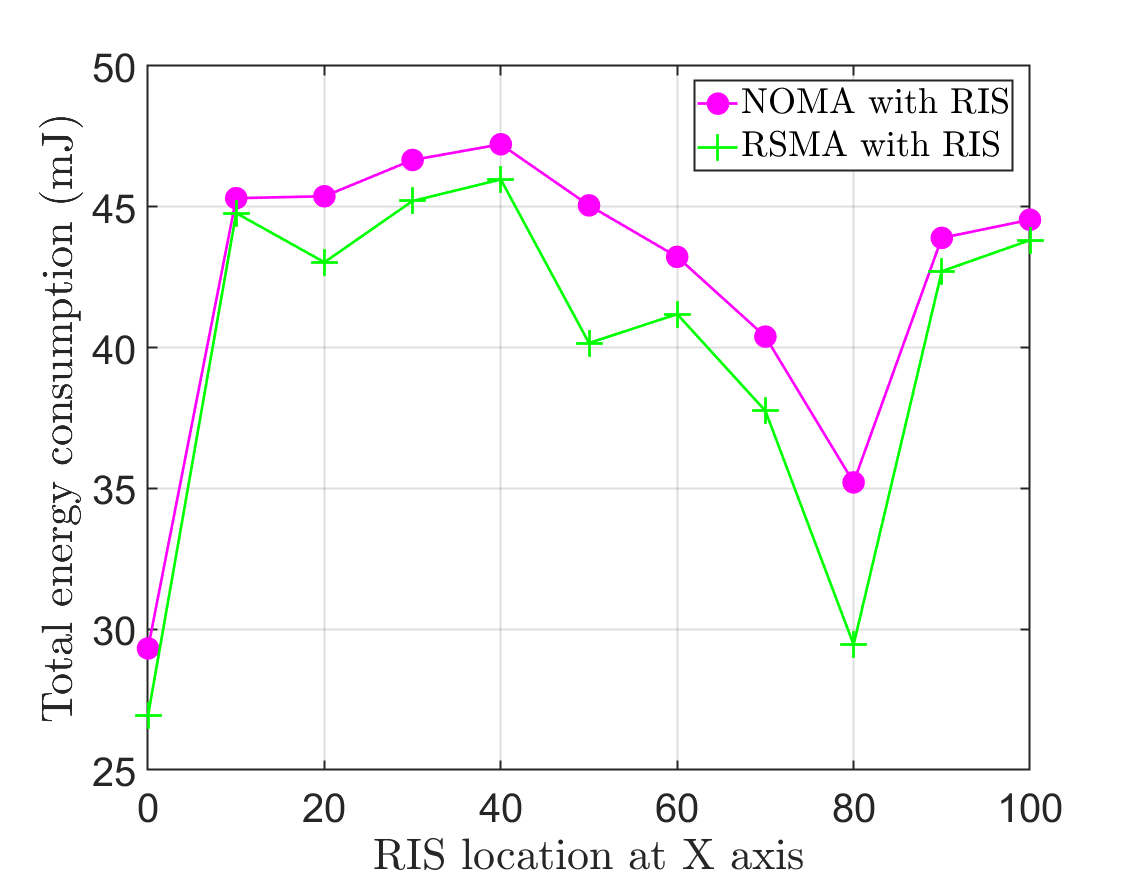}\label{Fig 6: ris loc rsma}}
    \caption{Energy consumption vs RIS location with $M=40$, $R_{th,1}=1$ bps/Hz and $R_{th,2}=3$ bps/Hz.}
\end{figure*}

Fig. \ref{Fig 5: RIS loc crsma} and \ref{Fig 6: ris loc rsma} present the impact of the RIS's location on the total energy consumption of four schemes that benefit from the RIS. The coordinate of RIS is set at $(X_{\rm RIS}, 10 \rm{m},0)$. It can be seen from Fig. \ref{Fig 5: RIS loc crsma} that the best location to place the RIS is near the BS, the near user, and the far user. This is due to the following three important reasons: 1) when the RIS is placed near the BS, RIS gets the opportunity to assist the BS directly during the transmission, and hence, the combined channel gains between BS $\rightarrow$ RIS $\rightarrow$ near user and BS $\rightarrow$ RIS $\rightarrow$ far user get enhanced, 2) when the RIS is placed close to the near user, it helps the BS transmission, in the first time slot, to the near UE as well as it helps the near user transmission to the far user in the second time slot to transmit with low power, which reduces the total energy consumption \cite{9586734}, 3) when the RIS is closer to the far user, the received SINR at the far user gets a supplementary boost in addition to the cooperation. 

\par Moving on to Fig. \ref{Fig 6: ris loc rsma} shows that in the case of general RSMA and NOMA (without cooperation), it is better to put the RIS beside the BS or close to the user with a weak channel gain. This is because the BS receives a boost in its transmission when the RIS is near to it. On the other hand, if the RIS is situated close to the weak user which has a low channel gain, the combined channel gain between BS, RIS, and weak user improves significantly, which makes the total energy consumption goes down.
\section{Conclusion}
\label{Conclusion}
In this paper, we studied a RIS-assisted C-RSMA
network. We aimed at jointly optimizing the precoding vectors at the BS, the common stream split, the relaying device power, the time slot allocation, and the phase shift matrix of RIS with the objective to minimize the total energy consumption. Due to the high coupling among the optimization variables, the formulated problem is a non-convex optimization problem. Hence, we decompose the original formulated problem into two sub-problems: sub-problem-1 and sub-problem-2. The first sub-problem is solved using an SCA-based low-complexity algorithm with a given phase shift matrix. Meanwhile, the second sub-problem is reformulated as a rank-one constrained optimization problem which is also solved using SCA. Utilizing the AO technique, the two sub-problems are solved alternately until convergence. Furthermore, to show the benefits of our proposed scheme, we compare our scheme with five other benchmark schemes, including the RSMA with RIS, NOMA with RIS, HD C-NOMA with RIS, HD C-RSMA without RIS, and HD C-NOMA without RIS.  Simulation results demonstrated that the amalgamation of RIS with HD C-RSMA enhances the system performance in terms of network energy consumption in comparison with the five benchmark schemes.
\bibliographystyle{IEEEtran}
\bibliography{IEEEabrv,Bibliography}

\begin{thebibliography}{10}
\providecommand{\url}[1]{#1}
\csname url@samestyle\endcsname
\providecommand{\newblock}{\relax}
\providecommand{\bibinfo}[2]{#2}
\providecommand{\BIBentrySTDinterwordspacing}{\spaceskip=0pt\relax}
\providecommand{\BIBentryALTinterwordstretchfactor}{4}
\providecommand{\BIBentryALTinterwordspacing}{\spaceskip=\fontdimen2\font plus
\BIBentryALTinterwordstretchfactor\fontdimen3\font minus
  \fontdimen4\font\relax}
\providecommand{\BIBforeignlanguage}[2]{{%
\expandafter\ifx\csname l@#1\endcsname\relax
\typeout{** WARNING: IEEEtran.bst: No hyphenation pattern has been}%
\typeout{** loaded for the language `#1'. Using the pattern for}%
\typeout{** the default language instead.}%
\else
\language=\csname l@#1\endcsname
\fi
#2}}
\providecommand{\BIBdecl}{\relax}
\BIBdecl

\bibitem{9693417}
Y.~Liu and other, ``{Evolution of NOMA Toward Next Generation Multiple Access
  (NGMA) for 6G},'' \emph{{IEEE} J. Sel. Areas Commun.}, vol.~40, no.~4, pp.
  1037--1071, Apr. 2022.

\bibitem{9831440}
Y.~Mao \emph{et~al.}, ``{Rate-Splitting Multiple Access: Fundamentals, Survey,
  and Future Research Trends},'' \emph{{IEEE} Commun. Surveys Tuts.}, pp. 1--1,
  2022.

\bibitem{mao2018rate}
Y.~Mao and \textit{et al.}, ``{Rate-splitting multiple access for downlink
  communication systems: bridging, generalizing, and outperforming SDMA and
  NOMA},'' \emph{EURASIP journal on wireless communications and networking},
  vol. 2018, no.~1, pp. 1--54, 2018.

\bibitem{mao2019rate1}
M.~Yijie and \textit{et al.}, ``{Rate-splitting for multi-antenna
  non-orthogonal unicast and multicast transmission: Spectral and energy
  efficiency analysis},'' \emph{{IEEE} Trans. Commun.}, vol.~67, no.~12, pp.
  8754--8770, 2019.

\bibitem{mao2020max}
Y.~Mao and \textit{et al.}, ``{Max-min fairness of K-user cooperative
  rate-splitting in MISO broadcast channel with user relaying},'' \emph{{IEEE}
  Trans. Wireless Commun.}, vol.~19, no.~10, pp. 6362--6376, 2020.

\bibitem{RIS_Open}
M.~Di~Renzo~\textit{et. al}, ``Reconfigurable intelligent surfaces vs.
  relaying: Differences, similarities, and performance comparison,'' \emph{IEEE
  Open Journal of the Communications Society}, vol.~1, pp. 798--807, June 2020.

\bibitem{9195473}
J.~Zhang \emph{et~al.}, ``{Energy and Spectral Efficiency Tradeoff via Rate
  Splitting and Common Beamforming Coordination in Multicell Networks},''
  \emph{{IEEE} Trans. Commun.}, vol.~68, no.~12, pp. 7719--7731, 2020.

\bibitem{9145200}
L.~Yin and B.~Clerckx, ``{Rate-Splitting Multiple Access for Multibeam
  Satellite Communications},'' in \emph{2020 IEEE International Conference on
  Communications Workshops (ICC Workshops)}, 2020, pp. 1--6.

\bibitem{7470942}
B.~Clerckx, H.~Joudeh, C.~Hao, M.~Dai, and B.~Rassouli, ``{Rate splitting for
  MIMO wireless networks: a promising PHY-layer strategy for LTE evolution},''
  \emph{{IEEE} Commun. Mag.}, vol.~54, no.~5, pp. 98--105, 2016.

\bibitem{7555358}
H.~Joudeh and B.~Clerckx, ``{Sum-Rate Maximization for Linearly Precoded
  Downlink Multiuser MISO Systems With Partial CSIT: A Rate-Splitting
  Approach},'' \emph{{IEEE} Trans. Commun.}, vol.~64, no.~11, pp. 4847--4861,
  2016.

\bibitem{8846761}
J.~Zhang \emph{et~al.}, ``{Cooperative Rate Splitting for MISO Broadcast
  Channel With User Relaying, and Performance Benefits Over Cooperative
  NOMA},'' \emph{{IEEE} Signal Process. Lett.}, vol.~26, no.~11, pp.
  1678--1682, 2019.

\bibitem{9123680}
Y.~Mao \emph{et~al.}, ``{Max-Min Fairness of \textit{K}-User Cooperative
  Rate-Splitting in MISO Broadcast Channel With User Relaying},'' \emph{{IEEE}
  Trans. Wireless Commun.}, vol.~19, no.~10, pp. 6362--6376, 2020.

\bibitem{9627180}
T.~Li \emph{et~al.}, ``{Full-Duplex Cooperative Rate-Splitting for Multigroup
  Multicast with SWIPT},'' \emph{{IEEE} Trans. Wireless Commun.}, pp. 1--1,
  2021.

\bibitem{9771468}
S.~Khisa \emph{et~al.}, ``{Full Duplex Cooperative Rate Splitting Multiple
  Access for a MISO Broadcast Channel with two Users},'' \emph{{IEEE} Commun.
  Lett.}, pp. 1--1, 2022.

\bibitem{9832618}
H.~Li \emph{et~al.}, ``{Rate-Splitting Multiple Access for 6G – Part III:
  Interplay with Reconfigurable Intelligent Surfaces},'' \emph{IEEE
  Communications Letters}, pp. 1--1, Jul. 2022.

\bibitem{huang2019reconfigurable}
H.~Chongwen \emph{et~al.}, ``{Reconfigurable intelligent surfaces for energy
  efficiency in wireless communication},'' \emph{{IEEE} Trans. Wireless
  Commun.}, vol.~18, no.~8, pp. 4157--4170, 2019.

\bibitem{yang2020energy}
Z.~Yang \emph{et~al.}, ``{Energy efficient rate splitting multiple access
  (RSMA) with reconfigurable intelligent surface},'' in \emph{2020 IEEE
  International Conference on Communications Workshops (ICC Workshops)}.\hskip
  1em plus 0.5em minus 0.4em\relax IEEE, 2020, pp. 1--6.

\bibitem{huang2018achievable}
C.~Huang \emph{et~al.}, ``{Achievable rate maximization by passive intelligent
  mirrors},'' in \emph{2018 IEEE International Conference on Acoustics, Speech
  and Signal Processing (ICASSP)}.\hskip 1em plus 0.5em minus 0.4em\relax IEEE,
  2018, pp. 3714--3718.

\bibitem{fang2022fully}
T.~Fang \emph{et~al.}, ``{Fully Connected Reconfigurable Intelligent Surface
  Aided Rate-Splitting Multiple Access for Multi-User Multi-Antenna
  Transmission},'' \emph{arXiv preprint arXiv:2201.07048}, 2022.

\bibitem{bansal2021rate}
A.~Bansal \emph{et~al.}, ``{Rate-Splitting Multiple Access for Intelligent
  Reflecting Surface aided Multi-User Communications},'' \emph{{IEEE} Trans.
  Veh. Technol.}, vol.~70, no.~9, pp. 9217--9229, 2021.

\bibitem{9759225}
K.~Weinberger, A.~A. Ahmad, A.~Sezgin, and A.~Zappone, ``Synergistic benefits
  in irs- and rs-enabled c-ran with energy-efficient clustering,'' \emph{IEEE
  Transactions on Wireless Communications}, pp. 1--1, 2022.

\bibitem{9393472}
A.~Bansal, K.~Singh, and C.-P. Li, ``Analysis of hierarchical rate splitting
  for intelligent reflecting surfaces-aided downlink multiuser miso
  communications,'' \emph{IEEE Open Journal of the Communications Society},
  vol.~2, pp. 785--798, 2021.

\bibitem{zhang2019cooperative}
J.~Zhang \emph{et~al.}, ``{Cooperative rate splitting for MISO broadcast
  channel with user relaying, and performance benefits over cooperative
  NOMA},'' \emph{{IEEE} Signal Process. Lett.}, vol.~26, no.~11, pp.
  1678--1682, 2019.

\bibitem{CSI_1}
L.~Wei \emph{et~al.}, ``{Channel Estimation for RIS-Empowered Multi-User MISO
  Wireless Communications},'' \emph{{IEEE} Trans. Commun.}, vol.~69, no.~6, pp.
  4144--4157, Jun. 2021.

\bibitem{gunduz2007opportunistic}
D.~Gunduz \emph{et~al.}, ``{Opportunistic cooperation by dynamic resource
  allocation},'' \emph{{IEEE} Trans. Wireless Commun.}, vol.~6, no.~4, pp.
  1446--1454, 2007.

\bibitem{amin2012opportunistic}
O.~Amin and L.~Lampe, ``{Opportunistic energy efficient cooperative
  communication},'' \emph{{IEEE} Wireless Commun. Lett.}, vol.~1, no.~5, pp.
  412--415, 2012.

\bibitem{AO_1}
W.~Ni \emph{et~al.}, ``{Resource Allocation for Multi-Cell IRS-Aided NOMA
  Networks},'' \emph{{IEEE} Trans. Wireless Commun.}, vol.~20, no.~7, pp.
  4253--4268, Jul. 2021.

\bibitem{AO_2}
M.~Elhattab \emph{et~al.}, ``{RIS-Assisted Joint Transmission in a Two-Cell
  Downlink NOMA Cellular System},'' \emph{{IEEE} J. Sel. Areas Commun.},
  vol.~40, no.~4, pp. 1270--1286, Apr. 2022.

\bibitem{mao2018energy}
M.~Yijie and \textit{et al.}, ``{Energy efficiency of rate-splitting multiple
  access, and performance benefits over SDMA and NOMA},'' in \emph{2018 15th
  International Symposium on Wireless Communication Systems (ISWCS)}.\hskip 1em
  plus 0.5em minus 0.4em\relax IEEE, 2018, pp. 1--5.

\bibitem{yang2020federated}
K.~Yang, T.~Jiang \emph{et~al.}, ``{Federated learning via over-the-air
  computation},'' \emph{{IEEE} Trans. Wireless Commun.}, vol.~19, no.~3, pp.
  2022--2035, 2020.

\bibitem{yu2021irs}
X.~Yu \emph{et~al.}, ``{IRS-assisted green communication systems: Provable
  convergence and robust optimization},'' \emph{{IEEE} Trans. Commun.},
  vol.~69, no.~9, pp. 6313--6329, 2021.

\bibitem{polik2010interior}
I.~P{\'o}lik and T.~Terlaky, ``{Interior point methods for nonlinear
  optimization},'' in \emph{Nonlinear optimization}.\hskip 1em plus 0.5em minus
  0.4em\relax Springer, 2010, pp. 215--276.

\bibitem{9145189}
Z.~Yang, J.~Shi, Z.~Li, M.~Chen, W.~Xu, and M.~Shikh-Bahaei, ``Energy efficient
  rate splitting multiple access (rsma) with reconfigurable intelligent
  surface,'' in \emph{2020 IEEE International Conference on Communications
  Workshops (ICC Workshops)}, 2020, pp. 1--6.

\bibitem{9197675}
F.~Fang \emph{et~al.}, ``{Energy-Efficient Design of IRS-NOMA Networks},''
  \emph{{IEEE} Trans. Veh. Technol.}, vol.~69, no.~11, pp. 14\,088--14\,092,
  2020.

\bibitem{9586734}
M.~Elhattab \emph{et~al.}, ``Reconfigurable intelligent surface enabled
  full-duplex/half-duplex cooperative non-orthogonal multiple access,''
  \emph{{IEEE} Trans. Wireless Commun.}, vol.~21, no.~5, pp. 3349--3364, 2022.

\bibitem{7117391}
Z.~Ding, M.~Peng, and H.~V. Poor, ``Cooperative non-orthogonal multiple access
  in 5g systems,'' \emph{IEEE Communications Letters}, vol.~19, no.~8, pp.
  1462--1465, 2015.

\end{thebibliography}
\vfill

\end{document}